\documentclass[superscriptaddress,notitlepage,longbibliography]{revtex4-1}
\pdfoutput=1
\usepackage{amsmath}
\usepackage{amssymb}
\usepackage{amsthm}
\usepackage{graphicx}
\usepackage{subfigure}
\usepackage[T1]{fontenc}
\usepackage[utf8]{inputenc}

\usepackage{array}
\usepackage{float}
\usepackage[table, usenames, dvipsnames]{xcolor}
\usepackage{color}
\newcolumntype{P}[1]{>{\centering\arraybackslash}p{#1}}

\usepackage{pifont}
\newcommand{\cmark}{\ding{51}}%
\newcommand{\xmark}{\ding{55}}%

\usepackage[colorlinks=true,urlcolor=blue,citecolor=blue]{hyperref}

\makeatletter
\@ifundefined{textcolor}{}
{%
 \definecolor{BLACK}{gray}{0}
 \definecolor{WHITE}{gray}{1}
 \definecolor{RED}{rgb}{1,0,0}
 \definecolor{GREEN}{rgb}{0,.4,0}
 \definecolor{BLUE}{rgb}{0,0,1}
 \definecolor{CYAN}{cmyk}{1,0,0,0}
 \definecolor{MAGENTA}{cmyk}{0,1,0,0}
 \definecolor{YELLOW}{cmyk}{0,0,1,0}
 }

\makeatother

\begin{abstract}
Open source software is becoming crucial in the design and testing of quantum algorithms.
Many of the tools are backed by major commercial vendors with the goal to make it easier to develop quantum software: this mirrors how well-funded open machine learning frameworks enabled the development of complex models and their execution on equally complex hardware.
We review a wide range of open source software for quantum computing, covering all stages of the quantum toolchain from quantum hardware interfaces through quantum compilers to implementations of quantum algorithms, as well as all quantum computing paradigms, including quantum annealing, and discrete and continuous-variable gate-model quantum computing.
The evaluation of each project covers characteristics such as documentation, licence, the choice of programming language, compliance with norms of software engineering, and the culture of the project.
We find that while the diversity of projects is mesmerizing, only a few attract external developers and even many commercially backed frameworks have shortcomings in software engineering.
Based on these observations, we highlight the best practices that could foster a more active community around quantum computing software that welcomes newcomers to the field, but also ensures high-quality, well-documented code.
\end{abstract}

\begin{document}
\title{Open source software in quantum computing}

\author{Mark Fingerhuth\footnote{Corresponding author: markfingerhuth@gmail.com}}
\affiliation{ProteinQure Inc., Toronto, Canada}
\affiliation{University of KwaZulu-Natal, Durban, South Africa}
\author{Tomáš Babej}
\affiliation{ProteinQure Inc., Toronto, Canada}
\author{Peter Wittek}
\affiliation{Rotman School of Management, University of Toronto, Toronto, Canada}
\affiliation{Creative Destruction Lab, Toronto, Canada}
\affiliation{Vector Institute for Artificial Intelligence, Toronto, Canada}
\affiliation{Perimeter Institute for Theoretical Physics, Waterloo, Canada}
\maketitle

\section*{Introduction}
Source code has been developed and shared among enthusiasts since the early 1950s.
It took a more formal shape with the rise of proprietary software that intentionally hid the code.
To counter this development, Richard Stallman announced the GNU Project in 1983, and started the Free Software Foundation.
Among other objectives, the aim of the project has been to allow users to study and modify the source code of the software they use.
This in turn formalized the concept of collaborative development of software products.
The term ``free'' as in freedom of speech has had philosophical and political connotations, but the model of massively distributed code development was interesting on its own right.
Collaborative communities around open source projects started emerging in the 1980s, with the notable examples of the GNU Compiler Collection (GCC) or the Linux kernel.
Soon thereafter, the widespread access to internet enabled many developer communities to successfully thrive and coordinate their efforts.

In the late 1990s, the term ``open source'' was coined to reflect the development model alone, and it was soon made mainstream by comparing the difficulties of monolithic software engineering to this new model.
This latter model is referred to as the ``the cathedral'', with a rigid development structure that may or may not meet user expectations.
This contrasts to the ``the bazaar'' model of open source, where the user needs drive the development, often in a haphazard fashion~\cite{raymond1999cathedral}).
The majority of open source contributors were volunteers, whose motivation varied from intrinsic reasons (e.g., altruism or community identification) or extrinsic (e.g., career prospects)~\cite{hars2002working}.
Many paid programmers contribute to open source projects as part of their job, which is another clear indication that open source is a software engineering paradigm as well as a business model~\cite{fitzgerald2006transformation}.

Open source software is a natural fit to scientific thinking and advancements and scientists have long embraced it with the TeX typesetting system being a prime example.
More recently, commercial entities started backing or even taking a leading role in open source software in science.
An example is machine learning: fairly complex mathematical models must be tested and deployed on hardware that is difficult to program and use to its full potential, for instance, on graphical processing units.
By providing high-quality open source frameworks, such as TensorFlow~\cite{abadi2016tensorflow} or PyTorch~\cite{paszke2017pytorch}, the commercial entities attract the best developers towards their ecosystem.
Parallel to these developments, quantum computers started to leave the labs and commercial entities began to sell computing time on this new type of hardware.
Thus, with a shared scientific and commercial interest in quantum computing, a tapestry of motivations emerges why the open source model is attractive for developing and distributing software in this domain.

\pagebreak
Let us highlight some of these motivations:
\begin{itemize}
\item Reproducibility: it is a core tenet of science.
The difficulties of reproducing results are, in the vast majority of cases, not due to the authors having made mistakes or, even worse, forged data.
Rather, the lack of detail about methods, data, and code in a paper is often the greatest impediment~\cite{ioannidis2008repeatability}.
Sharing the source code alleviates this problem~\cite{stodden2016enhancing}.
\item Impact and publicity: this is crucial for both scientific and commercial endeavours.
There is evidence that sharing details of the code will increase the impact of the work~\cite{piwowar2007sharing}.
\item Building a community and ecosystem: this is where the commercial angle is critical, since vendors of quantum hardware attract unpaid developers and potential future employees.
Traditionally, consulting and integrating businesses sprouted around large open source projects~\cite{hars2002working}. The difficulty with quantum computing is the steep learning curve that needs to be overcome, and therefore it is in the best interest of quantum hardware companies to get more developers involved.
This also resembles why and how large machine learning frameworks are supported by commercial entities.
\item Gaining credit and increasing human capital.
Learning new skills is an important motivation in contributing to open source projects~\cite{hars2002working}, and mastering quantum computing may be perceived as a pathway to better career prospects.
\end{itemize}

The outcome of the varied motivations is a plethora of tools, libraries, and even languages that are hosted in open source repositories.
This is the \emph{raison d'être} of this survey: we would like to raise awareness of open source projects in quantum computing, give credit to contributors, attract new developers to the field, and highlight the best aspects of some projects.
To achieve this, we divide the projects into categories that correspond to different levels of the quantum software stack, compare the projects, and highlight best practices that lead to success and more recognition.
We include contemporary, maintained projects according to a set of well-defined criteria, which also means that we had to exclude some seminal works, such as \texttt{Quipper}~\cite{green2013introduction}, \texttt{libquantum}~\cite{butscher2003libquantum} and \texttt{Liquid}~\cite{liquid}, which are no longer actively developed.
An accompanying website (\url{https://qosf.org/}) will receive automated updates on the projects to ensure that our work will continue to serve as a reference well after the publication of this paper.
This website is hosted in an open source repository and we invite the community to join the effort and keep the information accurate and up-to-date.

\section*{Software projects in quantum computing}
Experimental quantum computing is still a relatively new discipline and comparable to the early days of classical computers in the 1950s.
Similar to the manual programming of a classical computer with punch cards or an assembler, today's quantum computers require the user to specify a quantum algorithm as a sequence of fundamental quantum logic gates.
Thus, implementing a quantum algorithm on actual quantum hardware requires several steps at different layers of abstraction.

\begin{figure}[!h]
\includegraphics[scale=0.27]{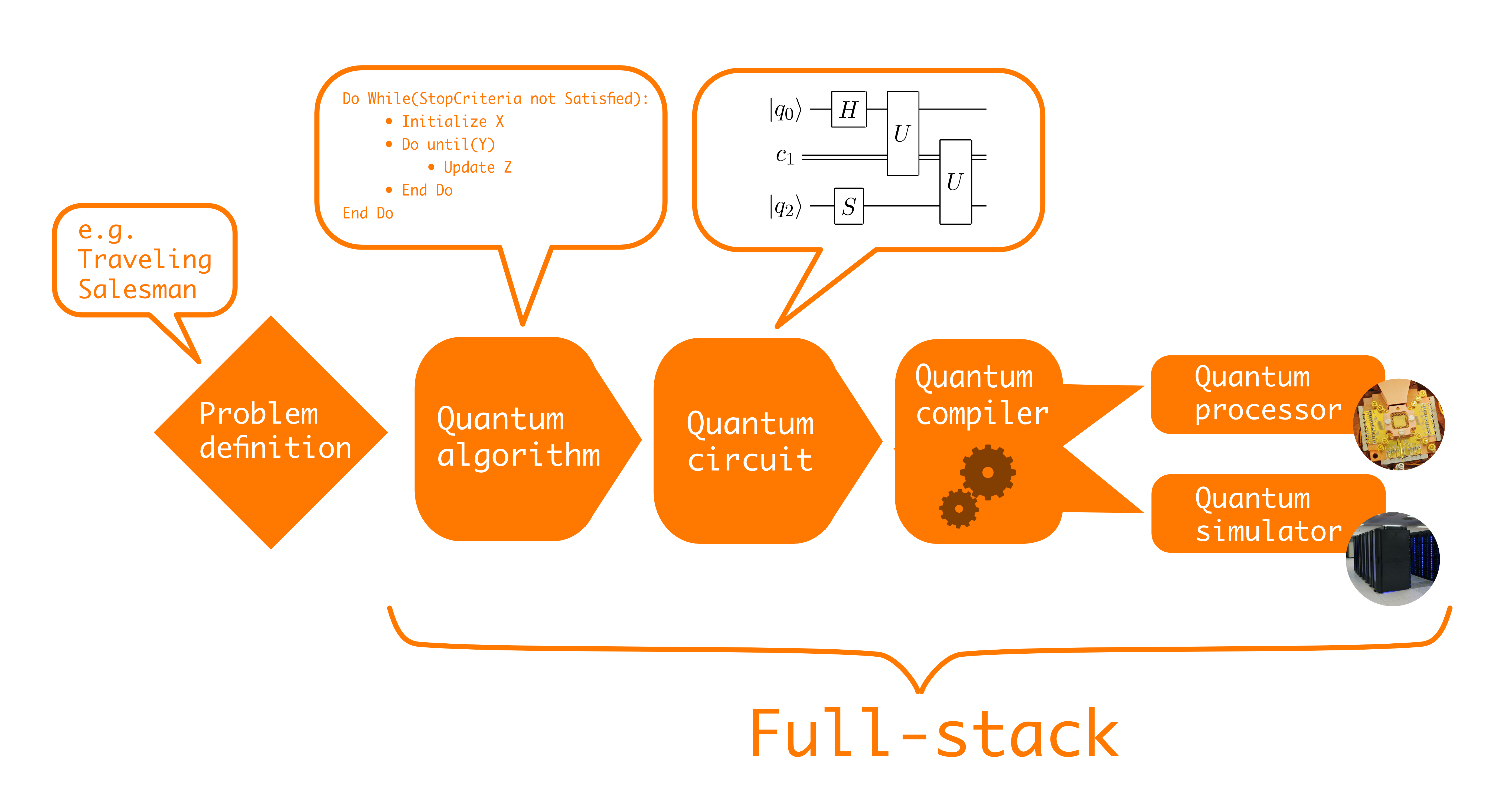}
    \caption{\label{img:universal_quantum_workflow}\textbf{Visualization of a typical quantum algorithm workflow on a gate-model quantum computer.} First, the problem is defined at a high-level and based on the nature of the problem a suitable quantum algorithm is chosen. Next, the quantum algorithm is expressed as a quantum circuit which in turn needs to be compiled to a specific quantum gate set. Finally, the quantum circuit is either executed on a quantum processor or simulated with a quantum computer simulator.}
\end{figure}

\begin{figure}[!h]
	\includegraphics[scale=0.27]{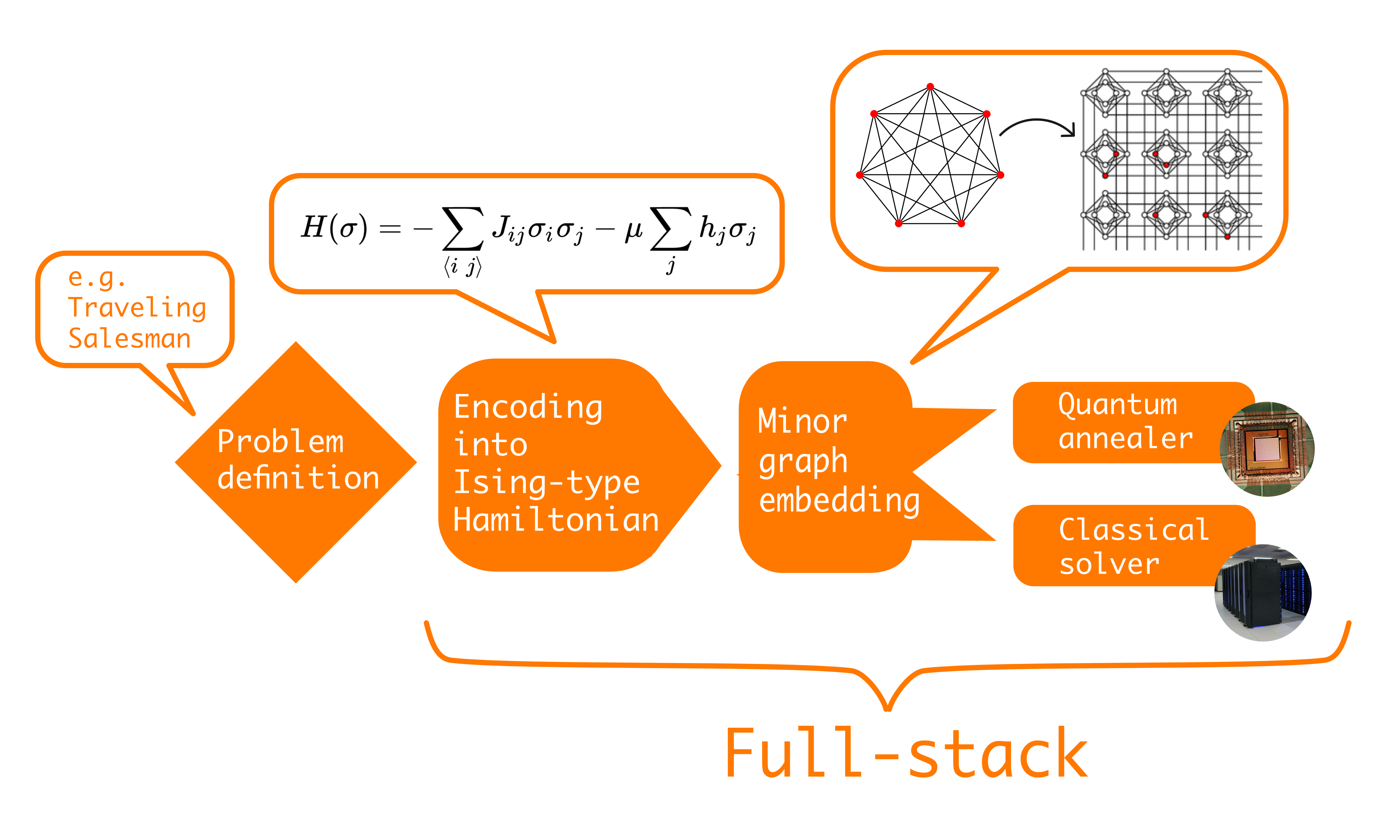}
    \caption{\label{img:quantum_annealing_workflow}\textbf{Visualization of a typical quantum algorithm workflow on a quantum annealer.} First, the problem is defined at a high-level and is then encoded into an Ising-type Hamiltonian which can be visualized as a graph. Next, via minor graph embedding the problem Hamiltonian needs to be embedded into the quantum hardware graph. Finally, either a quantum annealer or a classical solver is used to sample low-energy states corresponding to (near-)optimal solutions to the original problem.}
\end{figure}

To further complicate the picture, when we talk about quantum computing we talk about several different paradigms.
Some of these paradigms are barely abstracted away from the underlying physical implementation, which increases the difficulty of learning them for a computer scientist or a software engineer.
We define four paradigms:
\begin{enumerate}
    \item \textbf{Discrete variable gate-model quantum computing.}
	This is the generalization of digital computing where bits are replaced by qubits and logical transformations by a finite set of unitary gates that can approximate any arbitrary unitary operation.
	A classical digital circuit transforms bit strings to bit strings through logical operations, whereas a quantum circuit transforms a special probability distribution over bit strings -- the quantum state -- to another quantum state.
	Most quantum computing hardware companies focus on this model.
	For short, we refer to this model as the \emph{discrete gate model}.
    \item \textbf{Continuous variable gate-model quantum computing.}
	The qubits are replaced by qumodes, which take continuous values.
	Conceptually this paradigm is closer to the physics way of thinking about quantum mechanics, and quantum optics in particular.
	Most of the language that describes these circuits uses the terminology of quantum optics.
	We will refer to this model as the \emph{continuous gate model}.
    \item \textbf{Adiabatic quantum computation.}
	Quantum annealing devices exploit this model.
	At a high level, this paradigm uses a phenomenon from quantum physics known as the adiabatic theorem to find the global optimum of a discrete optimization problem.
	Recently, the actual physical devices that implement this paradigm have also been found useful in sampling a Boltzmann distribution.
	This paradigm does not have a direct classical analogue, and some understanding of statistical physics is recommended to work in this paradigm.
	This model will be referred to as \emph{quantum annealing}.
    \item \textbf{Quantum simulators.}
	These are application-specific quantum devices that are used, for instance, to study a particular model in quantum many-body physics.
	While this idea was the original motivation behind quantum computing~\cite{feynman1982simulating}, the field evolved significantly over the last three decades, and due to the lack of generality of this paradigm, we exclude it from our survey.
	Quantum simulators are not to be confused with simulations of quantum computation on classical computers.
\end{enumerate}
It remains challenging to understand what kind of problem can be solved efficiently by which paradigm and corresponding quantum algorithm.
A typical quantum algorithm workflow on a gate-model quantum computer is shown in Fig~\ref{img:universal_quantum_workflow}, whereas Fig~\ref{img:quantum_annealing_workflow} shows a typical workflow when using quantum annealing.
Both start with a high-level problem definition such as e.g. `solve the Travelling Salesman Problem on graph $X$'.
The first step is to decide on a suitable quantum algorithm for the problem at hand.
We define a \emph{quantum algorithm} as a finite sequence of steps for solving a problem whereby each step can be executed on a quantum computer.
In the case of the Travelling Salesman Problem we face a discrete optimization problem.
Thus, the user can consider e.g. the quantum approximate optimization algorithm~\cite{qaoa} that was designed for noisy discrete gate model quantum computers or quantum annealing to find the optimal solution.

Some of the open source projects require the user to define the quantum circuit of gates manually that represents the chosen algorithm for the given problem definition and quantum computing paradigm.
Other projects add another level of abstraction by allowing the user to simply define the graph $X$ and a starting point $A$, which encapsulates the Travelling Salesman Problem, and then automatically generate the quantum circuit for the chosen algorithm.
Note, that we explicitly distinguish a quantum algorithm from a quantum circuit.
A \emph{quantum circuit} is a quantum algorithm implemented within the gate-model paradigm whereas our notion of a quantum algorithm also includes the quantum annealing protocol.

If the scale of the quantum system is still classically simulable, the resulting quantum circuit can be simulated directly with one of the available open source quantum computer simulators on a classical computer.
The terminology is confusing, since hardware-based quantum simulators form a quantum computing paradigm as we classified above.
Yet, we also often call classical numerical algorithms quantum simulators that model some quantum physical system of interest, for instance, a quantum many-body system.
To avoid confusion, we will always use the term \emph{quantum computer simulator} to reflect that a quantum algorithm is simulated on classical hardware.

As opposed to a quantum computer simulator, the \emph{quantum processing unit} (QPU) is the actual quantum hardware representing one of the quantum computing paradigms.

QPUs and some simulators usually only implement a restricted set of quantum gates which requires compilation of the quantum circuit.
\emph{Compilation} connects the abstract quantum circuit description to the actual hardware or the simulator: it is the process of mapping the quantum gate set $G$ in a quantum circuit $C$ to a different quantum gate set $G^*$ resulting in a new quantum circuit $C^*$.
As an intuitive example, many quantum circuits use two-qubit gates between arbitrary pairs of qubits, even though those qubits might not be physically connected on the quantum processor.
Hence, the quantum compiler will swap qubits with each other until the required two qubits are neighbours, so the desired two-qubit gate can be implemented.
After applying the two-qubit gate we need to reverse the swaps to restore the original configuration.
The swaps require several extra gates.
For this reason, quantum circuits often increase in depth when being compiled.

The different steps in the quantum algorithm workflow outlined above mostly refer to the (continuous and discrete) gate models.
However, useful analogies can be made for the quantum annealing paradigm.
As shown in Fig~\ref{img:quantum_annealing_workflow}, having chosen quantum annealing as the quantum algorithm to tackle the Traveling Salesman Problem, the next step is to construct an Ising-type Hamiltonian that represents the problem at hand.
This is equivalent to constructing a discrete quantum circuit in the gate-model.
The actual QPU that performs the annealing seldom corresponds to the interaction pattern of the Hamiltonian.
For instance, the quantum annealing processors produced by D-Wave Systems currently have a particular graph topology -- the so called Chimera architecture -- that has four local and two remote connections for each qubit.
Thus, the previously generated problem graph must be mapped to the hardware graph by finding a minor graph embedding.
Finding the optimal graph minor is itself an NP-hard problem, which, in practice, requires the use of heuristic algorithms to find suitable embeddings~\cite{choi2008minor,choi2011minor}.
Finding a graph minor is analogous to quantum compilation and the size of the graph minor can be seen as the direct analogue to quantum circuit depth in the gate-model paradigm.
In-depth analyses of quantum annealing performance in Refs.~\cite{ronnow2014defining,venturelli2015quantum} have revealed a clear dependence between the quality of minor graph embeddings and QPU performance.
Lastly, the embedded graph can either be solved on a QPU or with a classical solver.
The latter is similar to using a quantum computer simulator in the gate-model paradigm.
When obtaining samples from a quantum annealer, it is common to further postprocess the results with classical algorithms to optimize solution quality~\cite{dwave_postprocessing}.
In both the gate-model and annealing paradigm, we define a \emph{full-stack library} as software that covers the creation, compilation / embedding, simulation and execution of quantum instructions as illustrated in Figs~\ref{img:universal_quantum_workflow} and \ref{img:quantum_annealing_workflow}.

Open source software in quantum computing covers all paradigms and all stages of expressing a quantum algorithm.
The software comes in diverse forms, implemented in different programming languages, each with their own vocabulary, or occasionally even defining a domain-specific programming language.
However, to provide a representative, but still useful study of quantum computing languages and libraries, we limited ourselves to projects that satisfy certain criteria.

\section*{Projects considered}

\begin{figure}[!ht]
\includegraphics[scale=0.17]{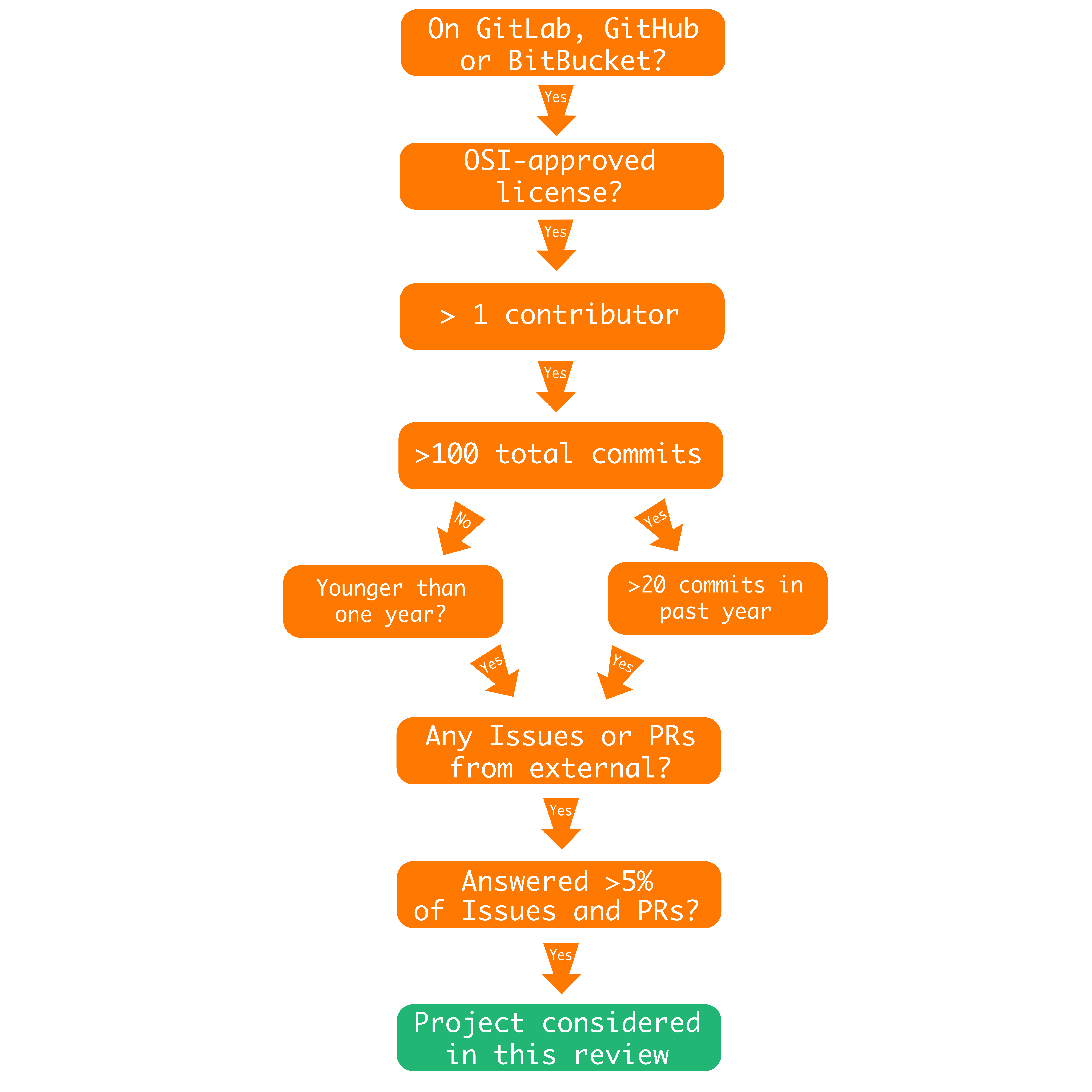}
    \caption{\label{img:selection_workflow}\textbf{Flow diagram with selection criteria.} Decision tree used to select quantum open source software projects for consideration in this study. The acronym PR stands for pull request which is a form of code contribution on software hosting websites.}
\end{figure}

In this section, we outline the criteria and explain our reasons for selecting them.
A concise overview of the selection process is depicted in Fig \ref{img:selection_workflow}.
In the second part of this section, we provide a brief outline for each of the selected projects.

\begin{itemize}
    \item \textbf{Open source community project}.
    In a highly collaborative field, such as quantum computing, it is fundamental for the researchers to cooperate.
    Empirically, there is evidence that open source community software projects survive and thrive on longer time scales than their closed source counterparts.
    Open source projects also seem to defy established theory by demonstrating increased productivity with an increasing amount of people~\cite{koch2004profiling}.
    However, high-quality open source software projects are often results of the dedicated work of individuals who have not made an effort to build developer communities around their projects, while still delivering high value to their (often anonymous) users~\cite{koch2005evolution}.
    For this reason, we do not restrict ourselves to projects that are being built by communities of developers in the literal sense of the word.
    Yet, the minimum requirement for a project to be considered is to have at least two contributors.
    Every project considered must have the potential to become a community collaborative effort.
        Hence, we require the project being developed under a version-control system (e.g. \texttt{git}) and hosted on one of the major software repository hosting sites for open source projects: GitHub, Gitlab and Bitbucket.
	This ensures basic visibility and the possibility of immediate contribution.

\item \textbf{Open Source Initiative (OSI)-approved licence}.
    An often misunderstood concept about open source projects is the issue of licencing.
    It might seem that all the projects on software repository hosting sites are considered open source simply because their source code is publicly available online.
    However, legally that is not the case since copyright laws apply by default~\cite{onlineChoosealicence}.
    An open source software project is one whose source code has been released under an open source licence, which gives the user a certain set of rights, namely, they allow the user to freely use, modify and share the software.
    The landscape of software licences is complex, with licences ranging from commercial to public domain.
    In this study we only consider projects released under one of the more than 60 licences that were approved by the Open Source Initiative (OSI)~\cite{osi}.

    \item \textbf{Project maturity}
    Code repository hosting sites are not only used for major projects, but often serve as a backup solution for prototypes, code stubs and toy examples.
    The line between the two is often hard to determine, because not all major projects are properly documented or attract outside interest.
    Hence, we use the number of commits as the proxy for determining whether a particular project is developed, enhanced and improved over a longer period of time.
    Since the frequency of developers' commits is partly influenced by personal style, we enforce the requirement of having at least 100 commits in the repository.

    \item \textbf{Activity}.
    An important characteristic of any open source software project is the existence of a maintainer, who is the main person responsible for the development and maintenance of the project.
    This role is initially fulfilled by the author of the project, but can eventually be passed on to a different community member.
    We only consider projects that have an active maintainer.
    A project is considered to have an active maintainer if it has a person actively developing the project, which includes either direct contributions from that person to the project's code base, or reviewing and accepting the changes proposed by external contributors.
    Both of these efforts manifests themselves as commits in the project's repository.
    Hence, we only consider projects having at least 20 commits in the past year.
    Those projects that are younger than a year are exempt from this condition.

\item \textbf{Outside interest}.
    Community open source projects do not thrive without contributions from users and fellow developers.
    Many repository hosting sites, including all that are considered in our survey, provide functionality to enable the community of users to interact with the maintainers and developers of the project.
    These interactions generally fall in two categories -- issues (which include questions and bug reports) and code contributions (often in the popular form of ``pull requests'').
    In this study, we concern ourselves with projects that have evidence of being useful to the wider community of quantum computing researchers.
    However, since not every user will contribute back to the project, we apply a fairly relaxed condition of at least having 1 code contribution or 1 bug report from a wider community of users. This excludes the core developers of the project as well as employees of the company hosting the project.
\end{itemize}

Let us briefly give an overview of all the open source projects considered in this review.
For a project to be considered, it had to fulfil the criteria outlined in Fig~\ref{img:selection_workflow}.
At this point, we would like to give an honourable mention to the quantum software project Q\#.
Most of it is licenced under custom licence terms, which are not recognized as an open source licence by the OSI, and therefore the project had to be excluded.
Table~\ref{tab:descriptions} lists all the selected projects and provides high-level information such as taglines, programming languages and supported operating systems.
Different projects cover different parts within the typical quantum algorithms workflows shown in Fig~\ref{img:universal_quantum_workflow} and \ref{img:quantum_annealing_workflow}.
Table~\ref{tab:feature_overview} illustrates this by clearly defining each project's range of applicability within the workflow.
We start at the bottom of the outlined technology stack with quantum computer simulators and compilers.

\begin{table}
    \footnotesize
    \begin{tabular}{p{2cm} p{9cm} p{2.1cm} p{1.6cm} p{1.6cm}}
		\multicolumn{5}{c}{Basic characteristics of included projects} \\
		\hline
		\hline
		Name & Tagline & Programming language & Licence  & Supported OS\\
		\hline
        \texttt{Cirq} &  Framework for creating, editing, and invoking Noisy Intermediate Scale Quantum (NISQ) circuits. & Python & Apache-2.0  & Windows, Mac, Linux \\
        \texttt{Cliffords.jl} & Efficient calculation of Clifford circuits in Julia. & Julia & MIT  & Windows, Mac, Linux \\
		\texttt{dimod} & Shared API for Ising/quadratic unconstrained binary optimization samplers. & Python & Apache-2.0  & Windows, Linux, Mac \\
        \texttt{dwave-system} & Basic API for easily incorporating the D-Wave system as a sampler in the D-Wave Ocean software stack. & Python & Apache-2.0  & Linux, Mac \\
        \texttt{FermiLib} & Open source software for analyzing fermionic quantum simulation algorithms. & Python & Apache-2.0  & Windows, Mac, Linux \\
        \texttt{Forest\newline (pyQuil \& Grove)} &  Simple yet powerful toolkit for writing hybrid quantum-classical programs. & Python & Apache-2.0  & Windows, Mac, Linux \\
        \texttt{OpenFermion} & The electronic structure package for quantum computers. & Python & Apache-2.0  & Windows, Mac, Linux \\
        \texttt{ProjectQ} & An open source software framework for quantum computing. & Python, C++ & Apache-2.0  & Windows, Mac, Linux \\
        \texttt{PyZX} & Python library for quantum circuit rewriting and optimisation using the ZX-calculus. & Python & GPL-3.0  & Windows, Mac, Linux \\
        \texttt{QGL.jl} & A performance orientated QGL compiler. & Julia & Apache-2.0  & Windows, Mac, Linux \\
        \texttt{Qbsolv} & Decomposing solver that finds a minimum value of a large quadratic unconstrained binary optimization problem by splitting it into pieces. & C & Apache-2.0  & Windows, Linux, Mac \\
        \texttt{Qiskit\newline Terra \& Aqua} & Quantum Information Science Kit for writing experiments, programs, and applications. & Python, C++ & Apache-2.0  & Windows, Mac, Linux \\
        \texttt{Qiskit Tutorials} & A collection of Jupyter notebooks using Qiskit. & Python & Apache-2.0  & Windows, Mac, Linux \\
        \texttt{Qiskit.js} & Quantum Information Science Kit for JavaScript. & JavaScript & Apache-2.0  & Windows, Mac, Linux \\
        \texttt{Qrack} & Comprehensive, GPU accelerated framework for developing universal virtual quantum processors. & C++ & GPL-3.0  & Linux, Mac \\
        \texttt{Quantum Fog} & Python tools for analyzing both classical and quantum Bayesian networks. & Python & BSD-3-Clause  & Windows, Mac, Linux \\
        \texttt{Quantum++} & A modern C++11 quantum computing library. & C++, Python & MIT  & Windows, Mac, Linux \\
        \texttt{Qubiter} & Python tools for reading, writing, compiling, simulating quantum computer circuits. & Python, C++ & BSD-3-Clause  & Windows, Mac, Linux \\
        \texttt{Quirk} & Drag-and-drop quantum circuit simulator for your browser to explore and understand small quantum circuits. & JavaScript & Apache-2.0  & Windows, Mac, Linux \\
        \texttt{reference-qvm} & A reference implementation for a Quantum Virtual Machine in Python. & Python & Apache-2.0  & Windows, Mac, Linux \\
        \texttt{ScaffCC} & Compilation, analysis and optimization framework for the Scaffold quantum programming language. & C++, Objective C, LLVM & BSD-2-Clause  & Linux, Mac \\
        \texttt{Strawberry Fields} & Full-stack library for designing, simulating, and optimizing continuous variable quantum optical circuits. & Python & Apache-2.0  & Windows, Mac, Linux \\
        \texttt{XACC} & eXtreme-scale Accelerator programming framework. & C++ & Eclipse\newline PL-1.0  & Windows, Mac, Linux \\
        \texttt{XACC VQE} & Variational quantum eigensolver built on XACC for distributed, and shared memory systems. & C++ & BSD-3-Clause  & Windows, Mac, Linux \\
    	\hline
\end{tabular}
\vspace{0.1cm}
    \caption{\label{tab:descriptions}\textbf{Overview of all projects considered in this review.} The table shows the tagline (description), the programming language(s) used, the licence and the supported operating systems (OS) for each project.}
\end{table}

\begin{table}
\begin{tabular}{ p{2.7cm} p{3.5cm} p{2.0cm} p{1.8cm} p{1.8cm} p{1.8cm} p{1.8cm} p{1.8cm}  }
		\multicolumn{8}{c}{Feature overview} \\
		\hline
		\hline
		Name & Quantum\newline computing paradigm & Quantum algorithms & Quantum circuits & Quantum compiler & Quantum computer simulator & QPU backend & Full-stack\\
		\hline
		\texttt{Cirq} & Discrete gate model & \cmark & \cmark & \cmark & \cmark & \xmark & \cmark \\
		\texttt{Cliffords.jl} & Discrete gate model & \xmark & \cmark & \xmark & \cmark & \xmark & \xmark \\
		\texttt{FermiLib} & Discrete gate model & \cmark & \xmark & \xmark & \xmark & \xmark & \xmark \\
        \texttt{Forest (pyQuil \& Grove)} & Discrete gate model & \cmark & \cmark & \cmark & \cmark & \cmark & \cmark \\
		\texttt{OpenFermion} & Discrete gate model & \cmark & \cmark & \xmark & \xmark & \xmark & \xmark \\
		\texttt{ProjectQ} & Discrete gate model & \cmark & \cmark & \cmark & \cmark & \cmark & \cmark \\
		\texttt{PyZX} & Discrete gate model & \xmark & \xmark & \cmark & \xmark & \xmark & \xmark \\
		\texttt{QGL.jl} & Discrete gate model & \xmark & \xmark & \cmark & \xmark & \xmark & \xmark \\
        \texttt{Qiskit Terra \& Aqua} & Discrete gate model & \cmark & \cmark & \cmark & \cmark & \cmark & \cmark \\
		\texttt{Qiskit Tutorials} & Discrete gate model & \cmark & \xmark & \xmark & \xmark & \xmark & \xmark \\
        \texttt{Qiskit.js} & Discrete gate model & \cmark & \cmark & \cmark & \cmark & \cmark & \cmark \\
		\texttt{Qrack} & Discrete gate model & \xmark & \cmark & \cmark & \cmark & \xmark & \xmark \\
		\texttt{Quantum Fog} & Discrete gate model & \cmark & \cmark & \xmark & \xmark & \xmark & \xmark \\
		\texttt{Quantum++} & Discrete gate model & \xmark & \cmark & \xmark & \cmark & \xmark & \xmark \\
		\texttt{Qubiter} & Discrete gate model & \cmark & \cmark & \cmark & \cmark & \cmark & \cmark \\
		\texttt{Quirk} & Discrete gate model & \cmark & \cmark & \xmark & \cmark & \xmark & \xmark \\
		\texttt{reference-qvm} & Discrete gate model & \xmark & \cmark & \xmark & \cmark & \xmark & \xmark \\
		\texttt{ScaffCC} & Discrete gate model & \xmark & \xmark & \cmark & \xmark & \xmark & \xmark \\
        \texttt{Strawberry Fields} & Continuous gate\newline model & \cmark & \cmark & \cmark & \cmark & \xmark & \cmark \\
		\texttt{XACC} & Discrete gate model & \cmark & \cmark & \cmark & \cmark & \cmark & \cmark \\
		\texttt{XACC VQE} & Discrete gate model & \cmark & \xmark & \xmark & \xmark & \xmark & \xmark \\
		\hline
		\hline
		Name & Hardware platform & Hamiltonian\newline generation & Minor\newline embedding & Post-\newline processing & Classical solver & QPU backend & Full-stack\\
		\hline
		\texttt{dimod} & Quantum annealing & \xmark & \cmark & \cmark & \cmark & \cmark & \xmark \\
		\texttt{dwave-system} & Quantum annealing & \xmark & \cmark & \cmark & \cmark & \cmark & \xmark \\
		\texttt{Qbsolv} & Quantum annealing & \xmark & \xmark & \xmark & \cmark & \cmark & \xmark \\
		\hline
\end{tabular}
\vspace{0.1cm}
    \caption{\label{tab:feature_overview}\textbf{Feature overview of selected projects.} Overview of the projects and how their features align with the typical quantum algorithms workflow shown in Figs~\ref{img:universal_quantum_workflow} and \ref{img:quantum_annealing_workflow}. Note, that the workflow is different in the quantum annealing paradigm as indicated by the reassigned column headings. Postprocessing is an additional feature used in quantum annealing to improve solution quality~\cite{dwave_postprocessing}. Data obtained in August 2018.}
\end{table}

\textbf{Quantum computing simulators.} \texttt{Quantum++} is a high-performance simulator written in C++ \cite{quantum++,qpp}.
Most quantum computer simulators only support two-dimensional qubit systems whereas this software library also supports the simulation of more general quantum processes.
\texttt{Qrack} is another C++ based simulator that comes with additional support for Graphics Processing Units (GPUs) \cite{qrack}.
The developers of \texttt{Qrack} put special emphasis on performance by supporting parallelization over multiple CPU or GPU cores.
A more educational and less performance-oriented quantum computer simulator is \texttt{Quirk} \cite{quirk}.
It is a JavaScript-based simulator that can simulate up to 16 qubits in a modern web browser.
\texttt{Quirk} provides a visual user experience by allowing beginners and experts to construct quantum circuits via simple drag-and-drop operations.
Next, Rigetti Computing, a hardware startup focused on superconducting circuits for the discrete gate model, has open sourced the project \texttt{reference-qvm} \cite{reference_qvm}.
This is a reference implementation of the Quantum Virtual Machine (QVM), synonymous with quantum computer simulator, used in their full-stack library \texttt{Forest}.
It is a purely Python-based simulator which is meant for rapid prototyping of quantum circuits.
So far, all mentioned quantum computer simulators simulate any quantum circuit until a certain depth.
This implies that these simulators support Clifford as well as non-Clifford quantum gates.
In contrast, the project \texttt{Cliffords.jl} restricts itself only to quantum gates from the Clifford group \cite{johnson2015demonstration,cliffords_jl}.
It is widely known that Clifford circuits can be simulated efficiently with a classical computer \cite{gottesman1998heisenberg} and
\texttt{Cliffords.jl} allows for fast and efficient calculations by making use of the tableau representation~\cite{aaronson2004improved} and it is written in the high-performance programming language Julia.

All quantum computer simulators so far are focused on the simulation of gate-model quantum computers.
Most of these simulators are used to develop and test quantum algorithms before implementing them on actual quantum chips or to verify results obtained from a QPU.
Analogously, the project \texttt{Qbsolv} is used to develop and verify the results obtained from quantum annealing devices \cite{qbsolv}.
Technically, it is not a quantum computer simulator as previously defined since it uses a classical algorithm unrelated to the physics of quantum annealing.
Yet, we are including it in this discussion because it is the closest analogue to a simulator for quantum annealing devices.
It is a C library that finds the minimum values of large quadratic unconstrained binary optimization (QUBO) problems.
To achieve this, the QUBO problem is first decomposed into smaller problems which are then solved individually using tabu search, a metaheuristic algorithm based on local neighbourhood search~\cite{glover1989tabu}.

\textbf{Quantum compilers.} Quantum computer simulators usually do not restrict the set of quantum gates (except \texttt{Cliffords.jl}) and allow two-qubit gates between any two simulated qubits.
However, when implementing quantum algorithms on actual hardware the circuits need to be compiled to the restricted topology of the particular quantum chip used for execution.
There are only a few standalone quantum compilers that satisfied our selection criteria.
Many quantum compilers are either absorbed into full-stack libraries, or they are proprietary and closed-source, developed by quantum hardware companies.
One of the few open source quantum compilers is \texttt{ScaffCC} which translates quantum circuits expressed in the Scaffold quantum programming language to quantum assembly format (QASM) \cite{javadiabhari2014scaffcc,scaff_cc}.
It also allows researchers to analyse the quantum circuit depth of quantum algorithm implementations on hypothetical future quantum chips.
Next, \texttt{QGL.jl} is a performance orientated quantum compiler for Quantum Gate Language (QGL) written in Julia \cite{qgl,ware2018experimental}.
Lastly, \texttt{PyZX} is a Python-based quantum compiler that uses the ZX calculus developed in Refs. \cite{coecke2008interacting} and \cite{coecke2011interacting} to rewrite and optimize quantum circuits \cite{pyzx}.

\textbf{Quantum full-stack libraries.} Several open source projects exist that move beyond isolated quantum computer simulation or quantum compilation and provide a full-stack approach to quantum computing as defined in Fig~\ref{img:universal_quantum_workflow}.

\texttt{ProjectQ}~\cite{haner2018software,steiger2018projectq,projectq}, \texttt{XACC}~\cite{xacc2018,xacc} and \texttt{Qubiter}~\cite{qubiter,tucci1999rudimentary} are the three quantum full-stack libraries that made all parts of their respective stacks available under open source licences.
\texttt{ProjectQ} was developed by researchers at ETH Zurich and is mostly written in Python \cite{haner2018software,steiger2018projectq,projectq}.
It allows the user to define, compile and simulate quantum circuits using an expressive syntax.
Additionally, \texttt{ProjectQ} can be used to interface with IBM's quantum processors through the cloud and support for other QPU backends is anticipated.
The \texttt{FermiLib} project completes the \texttt{ProjectQ} stack by providing a Python library to generate and manipulate fermionic Hamiltonians for quantum chemistry \cite{fermilib}.

Next, \texttt{XACC} is a C++ project that stands for eXtreme-scale ACCelerator and is an extensive quantum full-stack library developed with the support of the Oak Ridge National Laboratory~\cite{xacc2018,xacc}.
It is a software framework that allows the integration of QPUs into traditional high-performance computing workflows.
\texttt{XACC} has its own open source quantum compiler and supports execution on quantum chips from a wide range of quantum hardware companies as well as their respective simulators.
There is also an open source plugin that enables the use of a tensor network quantum virtual machine as a backend \cite{mccaskey2018validating,mccaskey2018hybrid}.
Finally, the project \texttt{XACC VQE} provides implementations of quantum chemistry algorithms for \texttt{XACC} \cite{xacc_vqe}.

The startup company Artiste-qb has open sourced their full-stack library \texttt{Qubiter} \cite{qubiter,tucci1999rudimentary}.
This library is mostly implemented in Python with some parts written in C++.
Additional to writing, compiling and simulating quantum circuits, \texttt{Qubiter} provides integration for the quantum processors of all major hardware providers.
Lastly, \texttt{Quantum Fog} is a separate open source project to generate and analyze quantum Bayesian networks \cite{tucci1995quantum}.
The authors plan to integrate it with \texttt{Qubiter} in the near future.

Rigetti Computing has open sourced most of its quantum full-stack library \texttt{Forest} \cite{rigetti_forest}.
\texttt{Forest} combines two separate open source projects, \texttt{pyQuil}~\cite{pyquil} and \texttt{Grove}~\cite{grove}.
\texttt{pyQuil} is an extensive Python library for the generation of Quil programs, where Quil is a quantum assembly language developed by Rigetti \cite{pyquil}.
It can be compiled using Rigetti's proprietary quantum compiler, which is not available under an open source licence.
Compiled Quil programs can then either be executed on their QPUs or simulated using Rigetti's Quantum Virtual Machine in the cloud or the reference implementation (\texttt{reference-qvm}) mentioned earlier~\cite{reference_qvm}.
\texttt{Grove} is the corresponding quantum algorithms library also written in Python~\cite{grove}.
It contains implementations of popular quantum algorithms such as the quantum approximate optimization algorithm~\cite{qaoa}, the variational quantum eigensolver~\cite{vqe} and the quantum Fourier transform.

IBM was first to provide public cloud access to their five qubit quantum processor in 2016. Since then, it has built a community of developers around their quantum software library \texttt{Qiskit} \cite{openQASM,qiskit}.
It is a full-stack library and consists of two separate projects, \texttt{Qiskit} \texttt{Terra} and \texttt{Aqua} \cite{qiskit_terra,qiskit_aqua}.
Similar to Rigetti's \texttt{Forest}, \texttt{Terra} is the base library that allows the user to define, compile and simulate quantum circuits, whereas \texttt{Aqua} is a collection of quantum algorithms implemented with \texttt{Terra}.
Furthermore, \texttt{Qiskit} provides the user with tools for quantum compilation and has a quantum computer simulator module as well as two freely accessible QPUs~\cite{ibm_qpus}.
Many of the algorithms in \texttt{Aqua} were outlined in Ref.~\cite{coles2018quantum} and there is an additional project, called \texttt{Qiskit Tutorials}, which contains many Jupyter notebooks with example code for programming in \texttt{Qiskit} \cite{qiskit_tutorials}.
In addition, \texttt{Qiskit.js} is a JavaScript clone of \texttt{Qiskit} providing the same functionality as discussed above \cite{qiskit_js}.

In contrast to quantum full-stack libraries that focus on the discrete gate model, \texttt{Strawberry Fields} is a Python-based library for the continuous gate model, developed by the startup Xanadu~\cite{killoran2018strawberry,strawberry_fields}.
It is based on the Blackbird quantum programming language and it is the only quantum software project built on top of a deep learning library: its computational backend for simulations is written in TensorFlow~\cite{abadi2016tensorflow}.
\texttt{Strawberry Fields}' repository contains example implementations of quantum algorithms, including quantum teleportation, boson sampling and several quantum machine learning algorithms.

Lastly, Google recently released their full-stack library \texttt{Cirq} \cite{cirq}.
It is written in Python and specifically aimed at the creation, compilation and execution of noisy intermediate scale quantum (NISQ~\cite{preskill2018quantum}) circuits.
\texttt{Cirq} has a parallelizable simulator backend that requires prior compilation to Google's preferred quantum hardware architecture.
The popular open source project \texttt{OpenFermion}, which generates fermionic Hamiltonians for quantum chemistry simulations provides a plugin to \texttt{Cirq} thereby completing the stack \cite{open_fermion,mcclean2017openfermion}.
Note, that \texttt{OpenFermion} is generally hardware-agnostic and can also integrate into the other quantum full-stack libraries such as \texttt{Forest} and \texttt{ProjectQ}.

\textbf{Quantum annealing.} There are several open source projects for quantum annealing with most projects being supported by D-Wave Systems.
We already mentioned \texttt{Qbsolv} together with the other quantum computer simulators, but would like to highlight two additional software projects.
The package \texttt{dimod} is a Python API for solving QUBO problems with various backends including D-Wave's quantum processors \cite{dimod}.
One of its unique contributions is the introduction of the binary quadratic model which unifies the Ising ($\pm 1$) and QUBO ($0/1$) formalisms.
Lastly, \texttt{dwave-system} is a GitHub project that provides a simple Python API interface to connect with the D-Wave Ocean software stack \cite{dwave-system}.
This allows the user to define QUBO problems as well as embed them onto a given quantum chip topology and optimize the minor embedding \cite{choi2008minor,choi2011minor}.
This project comes closest to a quantum full-stack library in the quantum annealing realm.

\section*{Evaluation}
\label{sec:evaluation}

Our aim for this section is to evaluate the surveyed projects based on a set of selected criteria, which emerge from best practices in open source software development.
This set of best practices is not set in stone.
As with any evolving system, they are subject to change, especially as new methods for software development and project management emerge, such as, for example, was the case when distributed version control systems largely replaced the centralized approach to software development.
Hence it is inevitable that there is some arbitrariness to our choices of criteria.
We anchor these choices in the literature and practice of software engineering to minimize arbitrariness.
For a thorough and practical introduction to building open source projects, please refer to Ref.~\cite{fogel2017ProducingOSS}, which provides the reader with an introduction on how to start projects and establish communities around them.

The structure of the section revolves around domain areas (or ``best practices'') in open source software development.
Background and reasoning for each domain area is concisely provided along with the criteria designed to evaluate the surveyed projects in this domain.
Furthermore, we provide quantitative and qualitative evaluation results in the form of summary tables.
As our data sources, we use static analysis of the source code, metadata from software repository hosting sites and in-depth analysis of the documentation to evaluate the included projects.

\textbf{Documentation.}
The first resource that enables users to get familiar with a new software project is its documentation.
Writing good documentation is a skill that is not necessarily aligned with the skill of being a good developer.
On the contrary, the authors and main developers of the project might be the wrong people to write the project's documentation, as they cannot easily take the perspective of a new user~\cite{fogel2017ProducingOSS}.
The duty is often fulfilled by the role of a technical writer in a commercial software development setting.
Documentation also gets outdated easily -- even software engineers in the commercial sphere often struggle to keep documentation up to date \cite{lethbridge2003HowSWEngUseDocs}.
Developers of open source projects have arguably even less incentive.
However, even bad or outdated documentation is better than none, as developers find incomplete or outdated documentation still useful \cite{lethbridge2003HowSWEngUseDocs}.
The guidelines described in the following paragraphs are a set of conventions that the community of open source developers have converged upon.

Good user documentation starts with a well written \texttt{README} file, which serves as the initial point of contact.
For most users, it is the first piece of documentation they encounter. Repository hosting websites like GitHub, Gitlab and Bitbucket even display the contents of the \texttt{README} file directly on the project's homepage.
As such, \texttt{README} files should provide a concise but thorough overview of the project, and serve as a hub to more detailed parts of the documentation.
As a minimum, the \texttt{README} file should clearly state the mission statement of the project, even if it may seem obvious from the project's name.
Next, it should communicate the capabilities (feature list) of the project on a high level, provide the build and setup instructions (or a link to the part of documentation that describes them) and list the major requirements of the project.
Adding an example of how to use the project or a screenshot in a practical setting can help users grasp what the project is offering.
In addition, it is always helpful to state the licence to clearly communicate the terms associated with using the code and mention main contributors / maintainers of the project to properly give credit where it is due.

Depending on their level of familiarity with the project, users come to consult documentation with two basic purposes in mind: newcomers seek information about how to actually use the project, while more experienced users might be interested in an explanation of more advanced features.
Good user level documentation accommodates for both of these expectations.
Hence, in our analysis, we differentiate between user documentation and detailed tutorials.
To get first time users familiar with the project, step-by-step tutorials are often used as they guide users through the process of using the library.
By \textit{user documentation}, we mean concise per-feature documentation.
This is the type of documentation mostly consulted by experienced users, since it provides an efficient and concise overview of a project's classes or functions and variables.
For example, a particular function and its arguments are described briefly and its usage is outlined with a small code example.
In many evaluated projects this is combined with automatically generated documentation as part of the source code documentation.
In general, missing pieces and deficiencies should also be mentioned explicitly, as it is hard for users to tell what features are present but not documented without consulting the source code.

Existing users need to be informed about recent changes and bug fixes in new releases such that they can adapt their usage of the code accordingly.
This is best done in the form of \textit{changelogs} that keep track of versioning and list new features as well as minor changes for each software release.
Changelogs should also give credit to developers that contributed to a particular release and ideally thank users who reported crucial bugs.

\begin{figure}[!ht]
\includegraphics[scale=0.8]{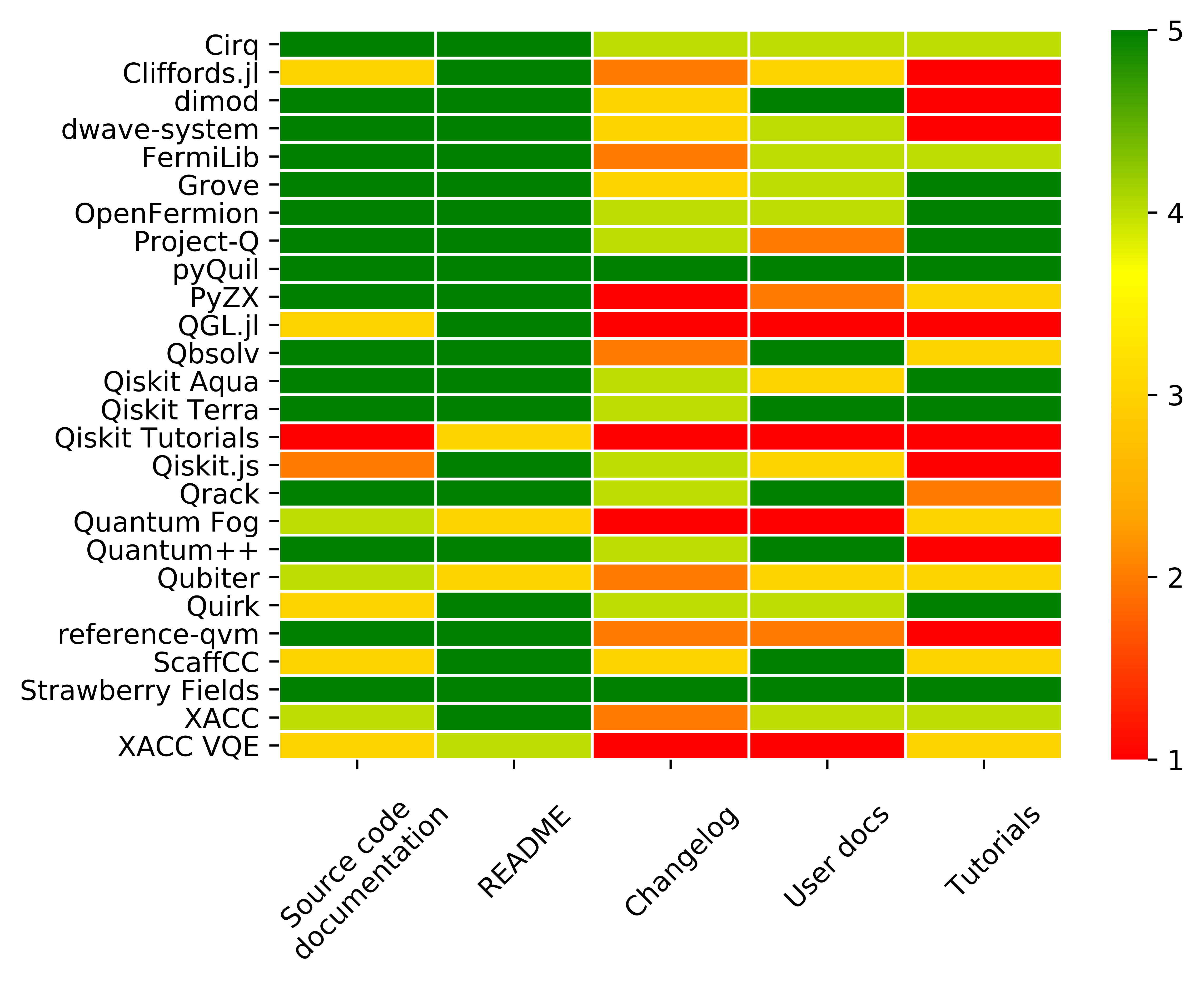}
    \caption{\label{img:heatmap}\textbf{Heatmap of documentation analysis results.} The heatmap shows the evaluation results for source code documentation, \texttt{README} files, changelogs, user documentation and tutorials on a scale from 1 (bad) to 5 (good). The evaluation rubrik used for scoring can be found in the Supporting Information in Table~\ref{tab:documentation_rubrik}. Data was obtained in August 2018.}
\end{figure}

Fig~\ref{img:heatmap} shows the detailed results of our qualitative documentation analysis in form of a colour coded heatmap with scores ranging from 1 (bad) to 5 (good).
The detailed rubrik used for scoring each of the five aspects can be found in the Supporting Information in Table~\ref{tab:documentation_rubrik}.

\textbf{User-centric discussion channels.}
It is understandable that users adopting a new project sometimes face the need to ask questions, even in presence of high-quality documentation.
Unlike commercial software, open source software is usually provided with no official support channels and hence this function is often voluntarily performed by the members of the wider user community of the project.
Yet, researchers in Ref.~\cite{hippelFreeUserToUser} have shown that despite the lack of direct funding, community driven support can outperform the support of commercially developed software in terms of quality.

However, the users cannot engage in the practice of answering each others' questions without the presence of a forum dedicated to this purpose.
Traditionally, this function is performed by user-centric mailing lists (or more historically, USENET groups), dedicated Q\&A sites, or forums.
Interactive communication channels, like IRC and Slack, Riot or Keybase are also useful, since the barrier to ask question is lowered in a chat-based interface.
Additionally, it also provides an easier, interactive debugging experience.
We have surveyed all projects with respect to their user-centric discussion channels and the results can be found in Table~\ref{tab:community_analysis}.

\begin{table}
	\begin{tabular}{p{3cm} P{1.5cm} P{1.5cm} P{2.5cm} P{1.7cm} P{1.7cm} P{2.0cm} P{2.1cm}  }
		\multicolumn{8}{c}{Community analysis} \\
		\hline
		\hline
        Project & Roadmap & Releases & Contributors & User-discussion channels & Developer-discussion channels & Public review processs & Community profile \\
		\hline
        \texttt{Cirq}              & \xmark & \cmark & 28 & Stack Exchange   & -   & E+I  & 4/7  \\
		\texttt{Cliffords.jl}      & \xmark & \cmark & 7  & -           & -        & E    & 3/7  \\
		\texttt{dimod}             & \xmark & \cmark & 11 & Forum       & -        & E+I  & 5/7  \\
		\texttt{dwave-system}      & \xmark & \cmark & 6  & Forum       & -        & E+I  & 4/7  \\
		\texttt{FermiLib}          & \xmark & \cmark & 10 & -           & -        & E+I  & 3/7  \\
        \texttt{Forest - Grove}    & \xmark & \cmark & 24 & Slack       & Slack    & E+I  & 3/7  \\
        \texttt{Forest - pyQuil}   & \xmark & \cmark & 46 & Slack       & Slack    & E+I  & 3/7  \\
		\texttt{OpenFermion}       & \xmark & \cmark & 26 & -           & -        & E+I  & 3/7  \\
		\texttt{ProjectQ}          & \xmark & \cmark & 10 & -           & -        & E+I  & 3/7  \\
		\texttt{PyZX}              & \xmark & \xmark & 3  & -           & -        & -    & 3/7  \\

		\texttt{QGL.jl}            & \xmark & \xmark & 3  & -           & -        & E+I  & 3/7  \\
		\texttt{Qbsolv}            & \xmark & \cmark & 18 & Forum       & -        & E+I  & 5/7  \\
		\texttt{Qiskit Aqua}       & \xmark & \cmark & 14 & Forum       & -        & E+I  & 7/7  \\
        \texttt{Qiskit Terra}      & \cmark & \cmark & 67 & Forum, Slack & Slack   & E+I  & 7/7  \\
		\texttt{Qiskit Tutorials}  & \xmark & \xmark & 37 & -           & -        & E+I  & 3/7  \\
		\texttt{Qiskit.js}         & \xmark & \cmark & 4  & Forum       & -        & E    & 7/7  \\
		\texttt{Qrack}             & \xmark & \cmark & 2  & -           & -        & E+I  & 3/7  \\
		\texttt{Quantum Fog}       & \xmark & \xmark & 2  & -           & -        & E    & 3/7  \\
		\texttt{Quantum++}         & \xmark & \cmark & 3  & Gitter      & -        & E    & 5/7  \\
		\texttt{Qubiter}           & \xmark & \xmark & 2  & -           & -        & E    & 3/7  \\
		\texttt{Quirk}             & \xmark & \cmark & 3  & -           & -        & E    & 4/7  \\
		\texttt{reference-qvm}     & \xmark & \cmark & 8  & -           & -        & E+I  & 3/7  \\
		\texttt{ScaffCC}           & \xmark & \cmark & 7  & -           & -        & E    & 3/7  \\
        \texttt{Strawberry Fields} & \xmark & \cmark& 5  & Slack        & Slack    & E+I  & 7/7  \\ 
		\texttt{XACC}              & \xmark & \xmark & 6  & -           & -        & E    & 4/7  \\
		\texttt{XACC VQE}          & \xmark & \xmark & 2  & -           & -        & E    & 3/7  \\
		\hline
	\end{tabular}
\vspace{0.1cm}
    \caption{\label{tab:community_analysis}\textbf{Evaluation results for the community analysis.} For each project, we indicate if a public development roadmap exists and if the software is published in form of releases. Additionally, we report the GitHub community profile score, the total number of contributors, the type of user- and developer-centric discussion channel and the type of public code review process -- specifically if it applies to internal (I) and/or external (E) contributors. Data obtained in August 2018.}
\end{table}

\textbf{Developer documentation.} 
The needs of project users are substantially different than the needs of the project's developers when it comes to documentation.
While users need to get familiar with outward facing parts of the project -- its API, requirements and the licence -- the developers of the project are interested in code documentation, exhaustive references and overall system architecture.

Appropriately addressing the expectations of developers -- such as maintaining high ratio of comments as a form of developer documentation -- increases the probability of obtaining an external contribution and thus converting a user into a member of the project's development community.
Furthermore, it also decreases the maintenance load on the existing developers~\cite{tenny1988}.

\textbf{Developer-centric discussion channels.}
External developer contributors in open source projects usually undergo an evolution, where they transition from passive users to external developers in multiple stages~\cite{nakakojiEvolutionPatterns}.
To encourage this transition, the existing developer community should be open enough such that passive users see a way how to participate in the development process.
This not only includes a process for accepting patches or pull requests, but also a public forum where developer and design discussions take place.
Developer-centric discussion channels conduct the majority of the design work in open source projects~\cite{mockus2002CaseStudies}.
In addition, they often host code reviews.
This is not only a fundamental practice for maintaining code quality within the project, but also serves as a process to increase code ownership and encourage knowledge transfer within the development team~\cite{bacchelli2013Expectations}.

However, a dedicated channel for developer discussions is not only useful for evolving contributors, but also beneficial for the existing, often distributed team of developers.
It has been shown that developers in distributed teams need to maintain awareness of each other~\cite{gutwinGroupAwareness}.
Developer mailing lists and more dynamic communication platforms, such as IRC channels or group chat applications are useful in maintaining awareness, even if not all discussions are code-related~\cite{guzzi2013Communication}.
Table~\ref{tab:community_analysis} shows the discussion platforms that are being used for these purposes by the analysed projects.

\textbf{Issue tracking system.}
When familiarizing themselves with a new piece of software, it is common for users to hit roadblocks.
Even seasoned users encounter bugs and rare problems when exploring more advanced parts of the code base.
Seeking help with their problems, users need a way to reach out to more experienced users, or even the project developers, and in the absence of better solutions end up using the project maintainer's email, if available.
This often leads to overloading the maintainer with repetitive questions and bug reports.

An issue tracking system helps avoid these problems.
It provides central overview of all the issues and bugs related to the project, and their status.
Some of the issues may turn out to be false positives, problems that are actually on the user side.
In these cases the issue tracking system serves as a knowledge base for the users experiencing the same problems.

Finally, we want to put emphasis on how a project deals with questions, issues and pull requests.
If an issue or pull request has seen no response from a core contributor (or an employee of the commercial entity backing the project) within 30 days we consider it as being ignored.
For this review we defined a \emph{core contributor} as a developer whose contributions sum to either 10~\% of all line additions or deletions or 15~\% of all commits.
We then define \textit{attention rate} as $1-I$ where $I$ is the fraction of ignored issues and pull requests with respect to the total number of issues and pull requests.
An ideal project never ignores any of its user or developer questions or contributions and would have an attention rate of $1.0$.
Lastly, we also compute the average time it takes a core contributor (or an employee of the company hosting the project) to respond to issues or pull requests.

\textbf{Roadmap.}
Well defined vision of the product is one of the best predictors of success in software development projects~\cite{werner2005}.
Since in open source projects the teams are not restrained to one geographical location, but distributed, the project's vision needs to be clearly communicated within the community.
A concise representation of the project's long term vision is often realized in form of the project's development roadmap.
Users benefit from the presence of the roadmap due to better clarity of the project's future development.
Developers are more encouraged to contribute, since they see their contributions in context.
Additionally, newcomers to the project can identify features that they have the skills and interest for.
Our community analysis in Table~\ref{tab:community_analysis} indicates which quantum software projects make use of development roadmaps.
Examples of successful open source projects outside of quantum computing that leverage (often community-defined) roadmaps are OpenStack \cite{onlineOpenStack} and OpenNebula \cite{onlineOpenNebula}.

\textbf{Outlining contribution process.}
Many open source projects have, over time, developed their own processes for accepting contributions.
This does not include only the technical requirements, but often contains conditions that a new contribution should adhere to, such as using the same code styling as the rest of the repository or making sure that the contribution is provided in a specific way (i.e. via pull request or a patch file submitted to the developer mailing list).

It is beneficial if these conditions and an overall description of the process are provided, such that expectations are set correctly and the risk of upsetting either side is minimized.
In Ref.~\cite{krogh2003community}, the authors concluded that newcomers to open source projects that follow established conventions in introducing themselves to the community and providing their contributions are more likely to be accepted.
Hence, it is safe to conclude that explicitly stating the conventions leads to faster and more successful contributions.
Many projects deal with this issue by providing a set of instructions in the \texttt{README} file, or even in a separate \texttt{CONTRIBUTING} file.
A good contributing file describes the expected way of interaction with other developers, outlines the expected process for reporting bugs and submitting patches / pull requests and also touches upon the governance structure of the project.

Since all projects considered in this review are hosted on GitHub, we chose one of GitHub's own metric, the ``community profile'', as a quantitative measure for quality of the community contribution process~\cite{github_community_profile1,github_community_profile2}.
The community profile evaluates seven aspects which are considered best practices for encouraging contributions by GitHub.
These include the existence of a short project description, a \texttt{README} file and a code of conduct which contains standards on how to engage in the community of developers.
Furthermore, a project should have a \texttt{CONTRIBUTING} file outlining how users can contribute to the project and a licence file that states how the code can be used.
Templates for issues and pull requests are also required for a complete community profile since they help streamline the process of issue tracking.
In our evaluation, we express the community profile as the fraction $X/7$ where $X$ is the amount of satisfied community profile requirements (see Table \ref{tab:community_analysis}).

\textbf{Usage of version control systems.}
The practice of versioning the code is an essential part of any software development team that consists of more than one person, especially in development teams that do not share the same geophysical location, as is often the case with open source community projects.
The particular type of the versioning system that is being used influences many aspects of the development process - the concurrent development of features, the review process as well as the likelihood of new contributions.

Based on their topology, version control systems are categorized as \textit{centralized} and \textit{decentralized}.
The examples of popular centralized version control systems are Subversion~\cite{zandstraSubversion} or CVS~\cite{berliner1990cvs}.
Decentralized version control systems nowadays are a more common industry practice, which is explained by the number of perceived advantages from the developer's point of view.
First of all, they treat all developers equally, as all developers have the ability to commit locally and hence maintain revisions. Additionally, they are often able to perform automated merges, simplify workflows for experimental branches and support work on the repository without internet connection~\cite{alwis2009WhyCentralized}.

\textbf{Licence.}
The source code of a software project is considered creative work and as such, in the absence of other arrangements, default copyright laws apply~\cite{onlineChoosealicence}.
Simply making the source code of the project publicly available, i.e. as publishing it on a code hosting site such as GitHub does not release the project into the public domain \cite{templeton1983publicdomain} and does not make it open source.
On the contrary, code that is made public without a licence is still considered proprietary and as such, not free to be used, shared or modified, even for non-commercial or research purposes~\cite{onlineGNU}.

Therefore the act of including a licence with the code -- formally referred to as releasing the code under a given licence -- is what is granting users and developers a set of rights to use, modify and share the project's source code.
As such, the presence of the licence under which the code repository is released is a fundamental part of the definition of an open source project.
In general, the open source software licences are divided into two groups -- so called permissive and copyleft.
Permissive licences tend to not restrict the users and developers, and allow the inclusion of the licenced code within commercial software.
Some licences include less severe restrictions, such as preserving attribution (i.e. The Apache 2.0 Licence \cite{onlineApachelicence}).
Copyleft licences, on the other hand, require the authors of the derivative works to redistribute their work under the same, or compatible copyleft licence.
The advantage of using a copyleft licence is in enforcing the open access even to the works that extend or otherwise build upon the original work.
However, that might be seen as restrictive, especially in commercially driven settings.
For a more thorough guide to the space of licencing of software, we recommend Ref.~\cite{morin2012QuickGuide}.
We provide an overview of the open source licences associated with the surveyed projects in Table~\ref{tab:descriptions}.

\textbf{Code readability.}
The readability of the code in open source projects is an important factor for maintainability and increases the probability of new developers contributing, since both current and new developers need to read parts of the existing codebase to contribute.
The act of reading code is considered the most time-consuming component in software maintenance~\cite{deimel1985reading}.
However, the notion of what properties make code readable easily gets subjective.
Suggestions like improving variable or method naming, code deduplication and simplifying loops, conditions and structures are common, universal improvements \cite{sedano2016Readability}. More subjective attributes like the indentation style or camel-casing need to preserved across the project for consistency.
Some projects, organizations and languages deal with this issue by imposing project-, company- or even language-wide code styling conventions (see for example Python's PEP8~\cite{onlinePEP8}).

To help with quantifying the notion of code readability, several metrics have been suggested~\cite{posnettAsimplermodel,aggrawal2002IntegratedMeasure}.
However, these are not widely used in practice, and code readability is often addressed as part of the code review process.
For the less empirical notion of code complexity, a popular metric is the cyclomatic complexity~\cite{mccabe1976complexity}.
It is a quantitative measure for the number of paths through the source code that are linearly independent.
Hence, a lower score (corresponding to lower complexity) is considered better, as it signifies a codebase that is less convoluted.
Cyclomatic complexity was only extracted from Python projects, which the majority of projects are, since the tool \texttt{radon} allows for easy extraction of this metric.
Unfortunately, other programming languages such as Julia, JavaScript or C++ provide no simple possibility for computing such metric.
The results for the Python projects are captured in Table~\ref{tab:static_analysis}.
Additionally, we conduct qualitative assessment of the source code comments in Fig \ref{img:heatmap} (see Table~\ref{tab:documentation_rubrik} in the Supporting Information for interpretation details).

\begin{table}[!h]
\small
	\begin{tabular}{p{3cm} P{1.5cm} P{1.5cm} P{1.5cm} P{1.5cm} P{1.5cm} P{1.8cm} P{1.8cm} P{2.1cm}  }
		\multicolumn{9}{c}{Static analysis} \\
		\hline
		\hline
        Name & Version control system & Issue tracking system & Issues/PRs & Attention rate & Average\newline response\newline time (days) & Test suite & Code coverage & Complexity \\
		\hline
		\texttt{dimod}             & Git & GitHub & 110/201 & 0.30 & 5.3  & \cmark & 94\%  & 2.96 \\
		\texttt{dwave-system}      & Git & GitHub & 54/72   & 0.24 & 8.2  & \cmark & 87\%  & 3.47 \\
		\texttt{Cirq}              & Git & GitHub & 448/686 & 0.54 & 2.6  & \cmark & 94\%  & 2.99 \\
		\texttt{Cliffords.jl}      & Git & GitHub & 6/12    & 0.33 & $<$1 & \cmark & -     & -    \\
		\texttt{dimod}             & Git & GitHub & 110/201 & 0.30 & 5.3  & \cmark & 94\%  & 2.96 \\
		\texttt{dwave-system}      & Git & GitHub & 54/72   & 0.24 & 8.2  & \cmark & 87\%  & 3.47 \\
		\texttt{FermiLib}          & Git & GitHub & 24/134  & 0.31 & $<$1 & \cmark & 99\%  & 2.43 \\
        \texttt{Forest - Grove}    & Git & GitHub & 53/130  & 0.51 & 17.7 & \cmark & 72\%  & 3.25 \\
        \texttt{Forest - pyQuil}   & Git & GitHub & 293/385 & 0.41 & 10.6 & \cmark & 88\%  & 2.65 \\
		\texttt{OpenFermion}       & Git & GitHub & 137/345 & 0.61 & 1.3  & \cmark & 100\% & 2.46 \\
		\texttt{ProjectQ}          & Git & GitHub & 84/198  & 0.75 & 4.0  & \cmark & 100\% & 4.02 \\
		\texttt{PyZX}              & Git & GitHub & 6/2     & 0.80 & $<$1 & \cmark & 51\%  & 4.42 \\
		\texttt{QGL.jl}            & Git & GitHub & 17/13   & 0.75 & 130.6& \cmark & -     & -    \\
		\texttt{Qbsolv}            & Git & GitHub & 50/85   & 0.17 & 22.2 & \cmark & 95\%  & -    \\
        \texttt{Qiskit Aqua}       & Git & GitHub & 43/141  & 0.20 & 1.8  & \cmark & 67\%   & 3.04 \\
        \texttt{Qiskit Terra}      & Git & GitHub & 526/713 & 0.11 & 16.0 & \cmark & 76\%  & 2.56 \\
		\texttt{Qiskit Tutorials}  & Git & GitHub & 94/274  & 0.40 & 8.6  & \xmark & -     & -    \\
        \texttt{Qiskit.js}         & Git & GitHub & 19/8    & 0.33 & 4.4  & \cmark & 66\%  & -    \\
		\texttt{Qrack}             & Git & GitHub & 7/78    & 0.07 & 8.7  & \cmark & 87\%  & -    \\
		\texttt{Quantum Fog}       & Git & GitHub & 17/1    & 1.00 & $<$1 & \xmark & 0\%   & 3.32 \\
		\texttt{Quantum++}         & Git & GitHub & 8/45    & 0.88 & $<$1 & \cmark & 72\%  & -    \\
		\texttt{Qubiter}           & Git & GitHub & 14/3    & 0.75 & $<$1 & \xmark & 0\%   & -    \\
		\texttt{Quirk}             & Git & GitHub & 286/131 & 0.96 & $<$1 & \cmark & -     & -    \\
		\texttt{reference-qvm}     & Git & GitHub & 6/14    & 0.44 & 75.6 & \cmark & 80\%  & 3.99 \\
		\texttt{ScaffCC}           & Git & GitHub & 15/11   & 0.18 & 10.1 & \cmark & -     & -    \\
        \texttt{Strawberry Fields} & Git & GitHub & 16/20   & 0.73 & 1.2  & \cmark & 97\%  & 2.70 \\
		\texttt{XACC}              & Git & GitHub & 65/14   & 0.65 & $<$1 & \cmark & -     & -    \\
		\texttt{XACC VQE}          & Git & GitHub & 22/4    & 0.33 & 8.8  & \cmark & -     & -    \\
		\hline
	\end{tabular}
    \vspace{0.1cm}
    \caption{\label{tab:static_analysis}\textbf{Evaluation results for the static analysis of each project and its source code.} We report the version control and issue tracking systems as well as the total number, attention rate and average response time for all open and closed issues and pull requests (PRs). Next, we analyze the existence of a test suite and report the resulting code coverage for most projects. Code complexity is only reported for projects written in Python since other languages do not allow for fast retrieval of this metric. Data obtained in August 2018.}
\end{table}

\textbf{Automated test suite.}
The benefits of automated software testing are widely accepted both in the academic sphere and between practitioners~\cite{peterseon2012Benefits}.
The two main approaches to automated testing are regression and unit testing, which are not mutually exclusive and the majority of projects employ both methods in their automated test suites.
Regression testing ensures that implemented functionality of the software, which is currently working, stays working after introducing a change.
Its goal is therefore to reduce the number of changes that ``break'' the existing functionality.
Such a change is also referred to as \textit{regression}.
Since long, regression testing has been identified as critical, and is one of the most widely used software testing strategies~\cite{onomo1998regressionIndustrial}.
Unit testing is focused on making sure that the module's boundaries (also referred to as its API) is respected.
Its goal is also to reduce the chance of introducing a breaking change, however, while regression testing is retrospective (making sure that regressions from the past will not reoccur), unit testing is considered prospective (trying to anticipate future breaking changes by testing a set of valid use cases)~\cite{fogel2017ProducingOSS}.

Code test coverage is long known to be one of the most important metrics in evaluating the quality of automated test suites.
For medium to large code bases, the code coverage has been shown to be highly correlated with software reliability~\cite{frate1995Correlations}.
In practice, many software development teams require maintaining code coverage of 85~\% or more in order to prevent introductions of defects into the software~\cite{williams2001CodeCoverageQuality}.

To evaluate the quality of the test suite of a particular surveyed project, we first evaluate whether the project has any test suite available.
If tests are present, we subsequently determine the corresponding code coverage across the whole code base, excluding maintenance and documentation files (such as tutorials and code examples, which might be still included in the repository, but not explicitly covered by tests).
To extract code coverage, we used the tools \texttt{pytest-cov}, \texttt{istanbul} and \texttt{gcov} for Python, JavaScript and C++ projects respectively.
The results of the evaluation are summarized in Table \ref{tab:static_analysis} (see columns ``Test suite'' and ``Code coverage'').
For most projects, running the test suite is rather straightforward but in some cases this is not the case.
In these cases, we skipped this part of the evaluation and left the corresponding fields blank.

\section*{Discussion}

To highlight the best practices and underline gaps in open source software in quantum computing, we interpret the evaluation results in the context of user, developer and community experience.
Additionally, we also talk about governance structure of open source development and the importance of open standards -- these aspects are external to the evaluation results, but they are important to the growth of the quantum computing community.

\subsection*{User experience}

\textbf{Documentation.}
All in all, producing up-to-date, extensive documentation is a hard task, and our survey confirms that.
Even if documentation efforts follow the open source collaborative approach~\cite{berglund2001OpenSourceDocs}, projects often still struggle with providing user-level documentation that covers the majority of features as well as providing step-by-step tutorials for newcomers.

By far the best results across all projects and categories were achieved in the \texttt{README} category.
All but three projects scored the maximum number of points for their \texttt{README} files.
Changelogs and user documentation, on the other hand, are major weaknesses of almost all projects.
We found that user documentation, if present, often only covers a small percentage of a project's functionalities.
This makes it difficult for new (and even experienced) users to understand the capabilities of the project at a quick glance.
Projects with outstanding user documentation include D-Wave's \texttt{Qbsolv} and \texttt{dimod} as well as the quantum computer simulator \texttt{Quantum++} and IBM's \texttt{Qiskit} library.
Despite the fact that almost all projects have a changelog, none, with the exception of \texttt{Strawberry Fields} and \texttt{pyQuil}, used them to give credit to contributors.
Most of the changelogs were sparse and contained very little information about recent changes.
This makes it hard for existing users to understand past development within the project and, most importantly, to adapt their own code to new releases.

Our results also show that most quantum software projects lack detailed code tutorials.
We would like to highlight \texttt{Forest} with its two subprojects \texttt{pyQuil} and \texttt{Grove} for its exceptionally detailed tutorials.
Their narrated tutorials, filled with illustrations, plots and code examples walk the user through various use cases of their software stack.
\texttt{Strawberry Fields}, developed by Xanadu, is another example.
Since it is the only full-stack library for the continuous gate model, the developers put emphasis on teaching the users basic concepts such as elementary gates, but still providing more complex tutorials for researchers experienced with the physics of this paradigm.
Having pioneered open source quantum computing, IBM's \texttt{Qiskit} achieves the same with respect to the theory of quantum computation in the discrete gate model.
All of the highlighted projects make it particularly easy for new users to get started.
Smaller projects tend to perform worse on this dimension since the preparation of tutorials is often time consuming and labour intensive.

\textbf{Issue tracking system.}
Essentially all the projects considered use issue tracking systems since they are an integral part of the software repository hosting websites.
Projects vary in their responsiveness to issues.
Out of 26 reviewed projects, only 15 had an average issue response time below one week.
In particular, \texttt{OpenFermion}, \texttt{Quirk} and \texttt{Strawberry Fields} are great examples for fast response times on issues and pull requests.
For example, even though the Google-backed project \texttt{Quirk} encounters a large number of issues and pull requests the core contributors respond, on average, in less than a day.
Such low response time encourages discussion and community building.
We would also like to highlight that most projects without commercial backing still manage good if not better attention rates and average response times.
\texttt{ProjectQ}, \texttt{Quantum++} and \texttt{Qubiter} are prime examples for high attention rates meaning that they take their issues and pull requests seriously.
This is remarkable since they have far less resources and are maintained by small groups of core contributors in their free time.

\textbf{User-centric discussion channels.}
We found that the majority of considered projects did not provide any communication platform for their users (see Table~\ref{tab:community_analysis}).
Only a few of the projects considered in the study have made the effort to provide their users with platforms for discussion and field support.
For the projects that offer user-centric discussion channels, dedicated online forums as well as the chat application Slack appear to be the primary choices.
For example, \texttt{Qiskit} users can seek help on the IBM Q Experience forum, \texttt{dimod} users can ask questions on D-Wave's Leap forum, whereas \texttt{pyQuil} users can interact with the wider community on the \texttt{Forest} Slack channel.
Our community analysis in Table~\ref{tab:community_analysis} clearly shows a lack of user support that goes beyond simply responding to issues and pull requests.
Giving users the ability to exchange ideas and help each other with problems requires dedicated discussion channels and we identify this as a major shortcoming in the field.

\textbf{Licence.}
Releasing the project's source code under an open source licence approved by the Open Source Initiative was one of our criteria for the inclusion of projects in this study (see Section 'Projects considered').
The projects vary in their choices for the licence.
The most popular licence was Apache-2.0, with 65~\% of the projects being released under its terms.
The other popular licences include BSD-3-Clause and MIT.
All of these three licences are considered permissive, and as such allow the derived software works to become proprietary (see Ref. \cite{wheeler_floss} for licencing implications of combining open source software).
Overall, only two projects were released under a copyleft GPL-3.0 licence, which is the type of licence originally conceived to counter the rise of proprietary software.
Choosing a more permissive licence is a sign that the community is open to commercial use of the existing infrastructure in general, while maintaining the advantages of open source development.

\subsection*{Developer and community experience}

\textbf{Usage of version control systems.}
As shown in Table~\ref{tab:static_analysis}, all the projects considered in this study used the distributed version control system \texttt{git} \cite{onlineGit,spinellis2012git,blischak2016quick}.
However, projects differ in how well they leverage the features that \texttt{git} offers.
The most common mode of operation for many projects is committing into one development branch (often called master).
This has the disadvantage of not providing stability of the project as new features are being developed.
We observe that the use of tags (i.e. commits that mark the new release versions) is fairly common among projects in our survey.
This is certainly a good practice that counterbalances the master branch acting as development branch in a project, since the tag serves as a long-term reference point to a particular state of the repository's code base.

\textbf{Roadmap.}
The vast majority of the projects analyzed did not provide a roadmap for future development of the project.
We believe this is an area for improvement, especially in the projects that have commercial backing where some of the decision process is hidden away from the wider community.
An exception is \texttt{Qiskit}, which provides a roadmap on \texttt{Qiskit Terra}'s wiki page.
Adopting a roadmap would be beneficial for smaller projects reviewed in this survey, as it communicates openness in the development process and could encourage additional developers to join the effort.

\textbf{Code readability.}
We find that on average, projects considered in this study achieve reasonable levels of code readability.
Due to prevalent usage of expressive programming languages such as Python, code complexity as measured by McCabe's approach~\cite{mccabe1976complexity} is fairly low.
Additionally, surveyed projects strive for good method/variable naming practices and include explanatory comments and docstrings.
This is evidenced by the prevalence of full-score for source code documentation in our quantitative analysis showcased in Fig \ref{img:heatmap}, which was achieved by 16 projects.
In particular, we would like to highlight projects exhibiting both high quality source code documentation and low code complexity, namely \texttt{Strawberry Fields}, \texttt{Qiskit Terra}, \texttt{Cirq}, \texttt{PyQuil}, \texttt{OpenFermion}, \texttt{FermiLib} and \texttt{dimod}.

\textbf{Automated test suite.} 
The popularity of automated test suites was confirmed in this study, as 23 projects out of the 26 studied employed a test suite.
However, the code coverage across projects varied widely, with an average of 75~\% and a standard deviation of 29~\%.
Yet, the median code coverage was found to be 87~\%, which is slightly above the industry-expected standard of 85~\%.

Many projects include small, automatically generated pictures (badges) showing the code coverage and status of the test suite (if all tests pass on the current branch).
We consider that a useful way of communicating a project's reliability since it also provides incentive to keep the test suite valid and code coverage high due to more immediate public pressure.

\textbf{Developer documentation.} 
While quite a few projects had reasonable internal documentation in form of docstrings and comments, we found high level approaches to projects, such as high level designs or system architecture diagrams, lacking.

\textbf{Developer-centric discussion channels.}
In general, projects considered in this study lack proper developer-centric communication channels.
To a certain degree this can be attributed to the size of the core development team, which is often very small (see number of contributors in Table~\ref{tab:community_analysis}).
In this case the overhead of a separate development-dedicated discussion channel might seem unnecessary.
However, the lack of a public discussion channel certainly does not encourage other users to contribute to the development process.
For larger, commercially backed, projects it seems that the majority of the design decisions happen behind the scenes, and even if the result of the process is open sourced, the wider community cannot directly participate in influencing the direction of the project.

The lack of developer-centric discussion channels does not inhibit code review in the projects, as this is often undertaken as part of the pull request process on GitHub.
Some projects have their core developers push their changes directly without any review approval from other team members.
A positive observation is that the majority of the surveyed projects conducts code review for external contributions as well as contributions from the core developer team (see column ``Public review process'' in Table~\ref{tab:community_analysis}).

\textbf{Outlining contribution process.}
In this study, projects largely ignored the need to set contribution guidelines.
This is reflected within GitHub's community profile metric and the relatively poor results in this category shown in Table~\ref{tab:community_analysis}.
\texttt{Strawberry Fields} and the three Qiskit projects -- \texttt{Terra}, \texttt{Aqua} and \texttt{Qiskit.js} -- are the only projects with maximum score on the community profile, since they have clear guidelines for future contributions.
Additionally, they have also created templates for their user's questions, issues and pull requests.

\textbf{Governance structure.}
The projects investigated in this survey do not follow any formal, public community governance structure.
This is a natural consequence of the general lack of public developer-centric discussion channels.
While some discussions happen as part of the process of reviewing pull requests from contributors, the majority of the development and design decisions are done offline.
This is also the case with the bigger projects in the survey (in terms of contributor count), as these are usually backed by commercial companies who presumably use internal meetings and discussion channels to drive decisions.
In order to drive community growth, we believe that projects would benefit from opening up the decision processes to the public, as doing so would share the sense of ownership of the project and encourage people to contribute.

Several models for governance have evolved in open source communities in the past.
Traditionally, open source projects start as single-person efforts.
However, as the project attracts more outside interest, growing its user and developer community requires a clear process for decision making.
Even though the projects often disregard formal approaches to governance, it has been established that such social systems need a certain form of governance structure to coordinate their efforts and scale~\cite{harrison1960}.
This is often realized throughout the lifetime of the project.
A thorough study of how governance structure evolves in an open source project can be found in Ref.~\cite{mahony2007}.

\textbf{Open standards.}
In recent decades, we have seen tremendous success being achieved with international collaboration pushing forward the capabilities of important technical projects such as the programming languages C/C++, or the networking protocol TCP/IP.
Even though multiple competing projects always exist, establishing an open standard ensures interoperability for the user, who, i.e. in the case of the C code, does not have to consider which compiler will compile the code, as long as the compiler satisfies the ANSI C standard.
Such standardization efforts reduce the burden on the user and increase portability of their code across multiple hardware architectures.

The field of quantum computing, and especially the open source software in this field is at the stage of rapid development.
Standardization has been neglected: all of the major players in the field are developing their own quantum computing domain-specific languages.
This need has been recognized by \texttt{ProjectQ} and \texttt{XACC}, which both aim to develop a common interface to universal quantum computing hardware providers.
Currently, however, \texttt{Project-Q} only supports the IBM quantum processors whereas \texttt{XACC} already offers support for QPUs from IBM, Rigetti and D-Wave.
In general, we identify the lack of open standardization in the field as a gap that will become more evident in the coming years.
Therefore we would like to encourage the wider community to establish discussions, working groups, or even open consortia to address these issues as their importance will only increase.

\section*{Conclusions}

This work started as a curated list of open source software projects in quantum computing.
We soon realized that curation requires well-defined criteria for evaluation and that this would have value to the community on its own.
Quantum computing is perplexing for newcomers, and the landscape of corresponding open source software is difficult to navigate even for seasoned quantum algorithm developers.
We hope that this survey gives credit to the pioneers and proves valuable to the growing community of quantum computing enthusiasts and experts.

The availability of these projects lowers the barrier to learn quantum computing: understanding, creating, and executing complicated mathematical models on esoteric hardware have become easier.
This reflects the same process that happened in machine learning, with solid open source frameworks supporting new and seasoned developers.

Yet, we identified several shortcomings, and projects with commercial support are not exempt from these findings.
Most projects lacked good documentation, making it difficult for new users to get started and to contribute in a meaningful manner.
Decision and design processes have largely been found to be conducted internally:
democratizing these could encourage the wider community to join the projects beyond the extent of sporadic contributions.
Several projects are slow at responding to user issues and pull requests.
Furthermore, we identified a lack of standardization in the field, where multiple players develop competing software platforms.
Maintenance and development overhead in the community would be greatly reduced both on the user and developer side if open standards were developed.
Finally, we identified a lack of stand-alone quantum compilers, since most compilers are either proprietary, closed-source or absorbed into quantum full-stack libraries.

As a paper on this topic quickly becomes outdated, we automated the extraction of the evaluation criteria and created a live website (\url{https://qosf.org/}) that updates the results as new releases of software appear.
This will ensure a lasting source of information on the field of quantum computing, which in itself welcomes contributions from the open source community.

\subsubsection*{Acknowledgements}
We would like to thank Josh Izaac (Xanadu), Alasdair G. Kergon (Red Hat), Alejandro Pozas-Kerstjens (ICFO-The Institute of Photonic Sciences), Nathan Killoran (Xanadu), Colin Lupton (Black Brane Systems), and Maria Schuld (Xanadu and University of KwaZulu-Natal) for discussions.

\pagebreak

\section*{Supporting information}

\begin{table}[!h]
    \begin{tabular}{| p{1cm} | >{\centering\arraybackslash}p{3cm} | >{\centering\arraybackslash}p{3cm} | >{\centering\arraybackslash}p{3cm} | >{\centering\arraybackslash}p{3cm} | >{\centering\arraybackslash}p{3cm} |}
		\hline
        \textbf{Score} & \textbf{Source code\newline documentation} & \textbf{\texttt{README}} & \textbf{Changelog} & \textbf{User documentation} & \textbf{Code tutorials}\\
		\hline
		1 & No comments or docstrings in the source code. & No \texttt{README} file exists. & No changelogs present. & No user documentation present. & No tutorials present.\\
		\hline
		2 & Some comments but no docstrings in the source code. & \texttt{README} file exists but only contains name and short description of project. & Only a list of commits without versioning. & Explanations/code examples that cover approx. 10\% of the total functionality. & Single poorly documented code example which combines several features into a workflow / algorithm.\\
		\hline
		3 & Plenty of comments and some docstrings in the source code. & Contains good and understandable project description. & List of new features for each version. & Explanations/code examples that cover approx. 40\% of the total functionality. & Multiple poorly documented code examples that combine several features.\\
		\hline
		4 & Well documented source code but source code documentation is not automatically generated. & Contains good project description and explanation of how to build/install the project. & List of new main features and minor changes for each version and each version is associated with corresponding commits. & Explanations/code examples that cover approx. 70\% of the total functionality. & Single narrated tutorial that guides the user through a use case of the project.\\
		\hline
		5 & Well documented source code and automatically generated source code documentation. & All of the above and a small code demo and/or contribution guidelines and description of licence and ideally badges for coverage, build status, etc. & Description of new features and minor changes and contributors and/or bug reporters are acknowledged. & Almost all features are documented with explanations and small code examples. & More than 3 narrated tutorials demonstrating use cases for the project.\\
		\hline
	\end{tabular}
\caption{\label{tab:documentation_rubrik}\textbf{Evaluation rubrik for code documentation.}}
\end{table}
\end{document}